\documentclass{nature}
\usepackage{graphicx}
\usepackage{caption}
\usepackage{amsmath}
\usepackage{bm}

\title{Neuromorphic signal processing with superconducting wirelet neurons}

\author{Khalil Harrabi$^{1,2}$, Leonardo R. Cadorim$^3$ \& Milorad V. Milo\v{s}evi\'c$^3$}

\begin{document}

\maketitle

\begin{affiliations}
  \item Physics Department, King Fahd University of Petroleum and Minerals (KFUPM), 31261 Dhahran, Saudi Arabia
 \item Interdisciplinary Research Center for  Advanced Quantum Computing, King Fahd University of Petroleum and Minerals (KFUPM), 31261 Dhahran, Saudi Arabia
 \item COMMIT, Department of Physics, University of Antwerp, Groenenborgerlaan 171, B-2020 Antwerp, Belgium
\end{affiliations}

\begin{abstract}
Neuromorphic computing aims to reproduce the energy efficiency and adaptability of biological intelligence in hardware. Superconducting devices are an attractive platform due to their ultra-low dissipation and fast switching dynamics. Here we employ a resistively shunted superconducting wirelet as a minimal artificial neuron for temporal neuromorphic computation. This simple architecture enables straightforward fabrication, electronic control, and high scalability. Through experiments and advanced simulations, we show that it exhibits spiking voltage dynamics driven by the interplay of resistive switching and relaxation, with threshold, firing frequency, and refractory time tunable through applied current, temperature, and shunt resistance. We demonstrate neural network operation by synaptically training temporal voltage signals generated by individual neurons. In this approach, trainable temporal weights act directly on the time-dependent wirelet-neuron responses rather than on static neuron outputs alone, allowing the computation to exploit the full temporal structure of the superconducting spikes. As an illustrative example, we apply this framework to handwritten digit recognition and show accurate classification using only three superconducting wirelet neurons. We further discuss on-chip training based on related gated wirelets as tunable synaptic elements, establishing shunted superconducting wirelets as scalable, energy-efficient building blocks for cryogenic artificial intelligence hardware that can be integrated with other emerging superconducting technologies.
\end{abstract}

Superconducting neurons are engineered to replicate the functionality of biological neurons using superconducting materials and circuits. These artificial neurons are characterized by key functional attributes, including spiking behavior, synaptic connectivity, and signal processing capabilities, which are essential for developing neuromorphic systems that emulate the brain's information-processing abilities. Central to the functionality of superconducting neurons is their ability to generate and propagate spikes, similar to the action potentials observed in biological neurons. In biological systems, neurons communicate via electrical spikes transmitted across synapses, leading to the activation or inhibition of other neurons within a network. Superconducting neurons typically replicate this spiking behavior using Josephson junctions, being superconducting devices that can easily switch between different states in response to an input signal. These junctions are configured to produce discrete voltage pulses analogous to spikes in biological neurons. These pulses can then be transmitted through superconducting circuits with minimal energy loss, enabling ultra-fast and efficient signal propagation. The spiking dynamics can be finely tuned by adjusting properties such as the critical current and junction capacitance, allowing for precise control over the neuron's behavior\cite{Goswami2020,Shainline2019}.

Beyond spiking behavior, superconducting neurons must establish synaptic connections with other neurons to form functional networks. In biological systems, synapses serve as junctions where signals are transmitted from one neuron to another, playing a critical role in learning and memory by modulating the strength of these connections. Synaptic functionality of superconducting neurons can also be implemented with Josephson junctions or other superconducting elements. These synapses can be engineered to modulate the strength and timing of signals, mimicking the plasticity observed in biological synapses. For example, synaptic weight can be adjusted by varying the current or magnetic flux applied to the Josephson junction, enabling the network to learn and adapt based on input signals. This plasticity is fundamental for neuromorphic systems designed to perform learning and memory tasks\cite{Segall2019,Shainline2018}.

Superconducting neurons are also adept at sophisticated signal processing, which is vital to perform complex computational tasks. In a network of superconducting neurons, each neuron integrates input signals from multiple sources, processes them according to its internal state, and generates an output spike if a certain threshold is reached. This mechanism is analogous to the integration-and-fire model of biological neurons, where inputs are summed, and an action potential is generated if the membrane potential surpasses a threshold. The high-speed operation of superconducting circuits allows these neurons to process signals at extremely fast rates, which is crucial for real-time applications such as artificial intelligence (AI) and quantum computing. Superconducting neurons can be coupled with superconducting qubits in a quantum processor, enabling integrated hybrid systems that combine the strengths of both quantum and neuromorphic computing. Furthermore, the low energy dissipation in superconducting circuits ensures that such computations can be performed with minimal power consumption, making superconducting neurons indispensable for large-scale networks\cite{Carnevale2021}.

However, the superconducting neurons based on Josephson junctions suffer from rather high complexity. Their circuit architectures, such as single-flux quantum (SFQ) and adiabatic quantum-flux-parametron (AQFP) systems, are readily used to implement neuronal functions like integration, inhibition, and temporal summation. However, these architectures require careful biasing schemes and precise magnetic flux control to prevent cross-talk and unintended coupling between neighboring circuits. Moreover, implementing synaptic plasticity in Josephson-based systems adds another layer of complexity. Superconducting synapses mostly rely on Josephson junctions or hybrid elements that can adjust synaptic weight via controlled magnetic fields or current biases, demanding delicate calibration and external feedback circuitry. When scaled up, the cumulative wiring, biasing networks, and thermal load further complicate the integration of thousands or millions of neurons. Consequently, while Josephson circuits offer high speed and energy efficiency, their intricate dynamics, sensitivity to fabrication tolerances, and stringent cryogenic requirements make superconducting neurons among the most technically demanding neuromorphic systems under development today.

Instead, in this work we bring forward a shunted superconducting filament, biased near its critical current, as the minimal yet elegant ``integrate-and-fire'' superconducting neuron. Superconducting nanowires were readily shown to behave as a biological neuron in Ref.~\citenum{toomey2019}, although there the firing threshold and recovery time were distributed across several coupled nonlinear systems, making it more of a hybrid electrothermal oscillator than a purely single-element neuron. Moreover, the neuron performance was reduced to just fire-or-not behavior, with subsequent electronic neural network simulated with LTspice software package\cite{ltspice}. Here we will focus on the functionality of a single superconducting filament as a neuron, and the integration of their temporal responses to the neuromorphic circuit where synaptic part also can be made of superconducting filaments, with electronically controlled synaptic weight. Moreover, we opt experimentally for micrometer-scale superconducting wires, as robust, controllable, and thermally stable neurons, capable of firing reliably at high rates and integrating cleanly with other superconducting circuits. Nanowires, while attractive for extreme miniaturization and lower power, suffer from instability, thermal noise, and fabrication sensitivity, making them less suitable for complex neuromorphic architectures — at least with current technology. We note however that all the concepts and neuromorphic strategies presented in this paper can be implemented on either nano- or micro-scale.

\section*{Neural properties of a superconducting wirelet}
A biological neuron displays a collection of characteristic behavior\cite{izhikevich2004,toomey2019,crotty2023} that any artificial analogue must emulate. Here we briefly characterize a shunted superconducting wirelet as a neural element and discuss how it mimics the expected biological behavior. 

\begin{figure*}[!t]
    \centering
    \includegraphics[width=\linewidth]{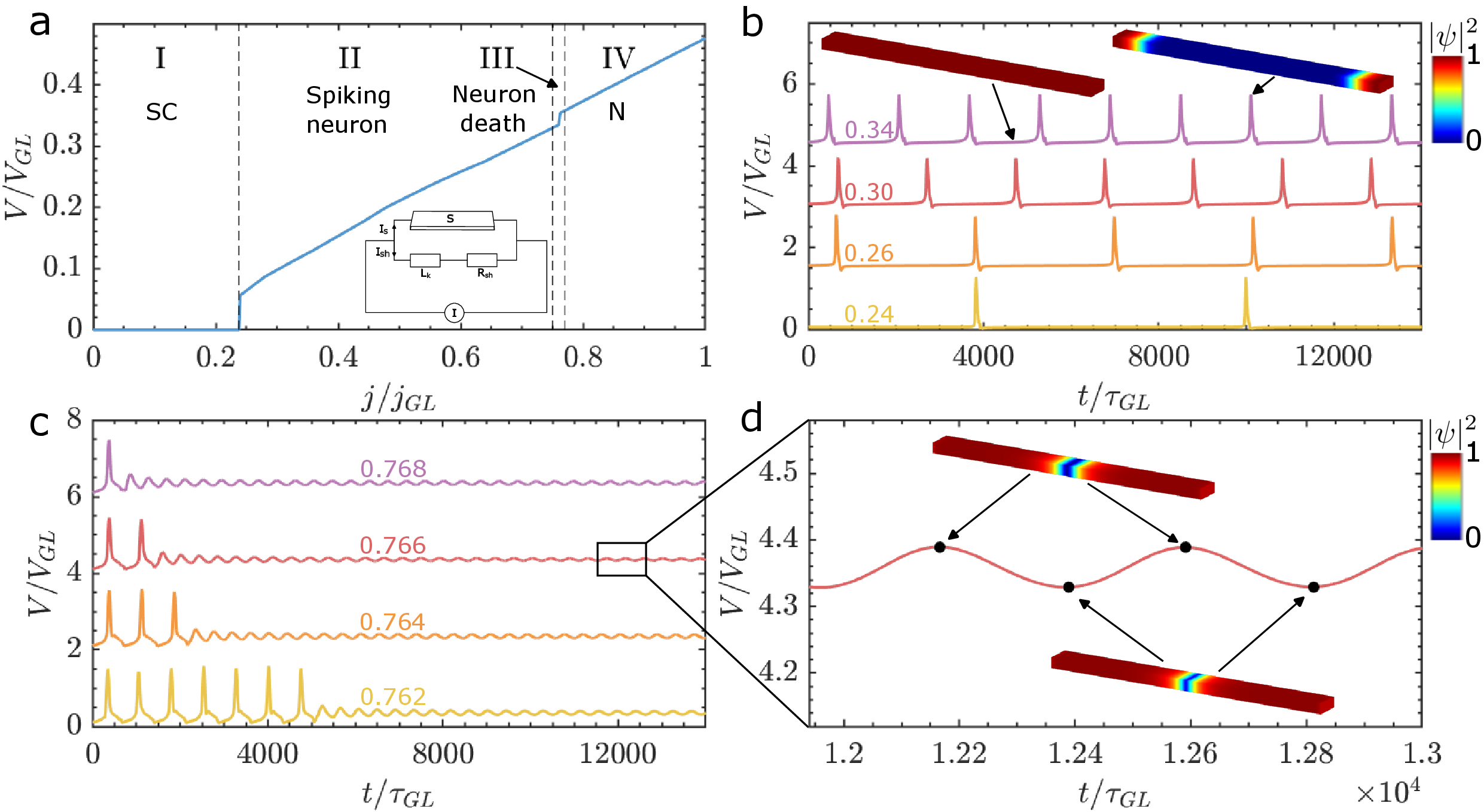}
    \caption{\textbf{Simulation insight into the resistive state of a shunted superconducting filament.} \textbf{a}, The calculated current-voltage characteristics of a shunted nanowire (width $10\xi$, schematic circuit in inset, with $L_k=500L_{GL}$ and $R_{sh}=0.5R_{GL}$; see details in Methods), with four main phases labeled. \textbf{b}, The spiking temporal behavior of the output voltage, for increased applied current (shown on each curve, in units of $j_{GL}$) within phase II. $V(t)$ curves are vertically displaced for visibility. Insets show the snapshot of the superconducting condensate in the filament before and during the voltage spike. \textbf{c}, Temporal behavior of the output voltage within phase III, showing the transformation of spiking behavior into damped oscillations corresponding to the continuous phase-slip state. \textbf{d}, Zoom-in on the phase-slip state from panel c), with visualization of its Cooper-pair density profile.} 
    \label{fig:fig1}
\end{figure*}

The first fundamental property of biological neurons is the \textit{threshold response}, i.e. a minimum intensity of the input signal below which the neuron would not fire. As seen from the calculated current-voltage (I-V) characteristics of a shunted superconducting wire in Fig.~\ref{fig:fig1}a, this behavior is promptly replicated by our artificial neuron, with the intensity threshold being identified with the critical current $I_c$ that separates the dissipationless superconducting state from the resistive state, where voltage spikes appear due to the periodic creation of hot spots\cite{Skocpol1974, Kerman2009, Zotova2012} (see insets of Fig.~\ref{fig:fig1}b) coupled to the current diversion to the shunt resistor.

Further, the real neurons are able to respond (fire) to two consecutive input signals only if the time interval between them is not shorter than a characteristic time called the \textit{refractory period}. In our element, this period is bound by the so-called delay time\cite{harrabi2024}, which is the time required for an applied current (above $I_c$) to disturb the superconducting state and create a hot spot. Since the delay time decreases with applied current, we note that the refractory period of our neuron is also current dependent. At the same time, as seen in the simulated $V(t)$ curves in Fig.~\ref{fig:fig1}b, our element presents the firing frequency that increases with the input (current) intensity. This corresponds to \textit{class-I behavior} of biological neurons.

Finally, our shunted superconducting wirelet can also replicate the death of a biological neuron, i.e. seizure of its ability to fire. This behavior is shown in Fig.~\ref{fig:fig1}c-d, where for a sufficiently large applied current our neuron no longer exhibits spikes in the output voltage. As detailed by the simulations, in such a scenario, the system reaches a dynamical equilibrium where, instead of the periodic hot-spot creation/relaxation process, we observe dynamic stabilization of a phase-slip line\cite{Langer1967, McCumber1970, Arutyunov2008} (see insets of Fig.~\ref{fig:fig1}d and Supplementary Video 1), with the current distribution between the superconductor and the shunt no longer varying in time. This behavior corresponds to the `dead mode' of a neuron\cite{schegolev2023} and has not been discussed previously in the context of shunted superconducting nanowire elements.

\section*{Experimental realization of a spiking neuron}
\begin{figure*}[!t]
    \centering
    \includegraphics[width=\linewidth]{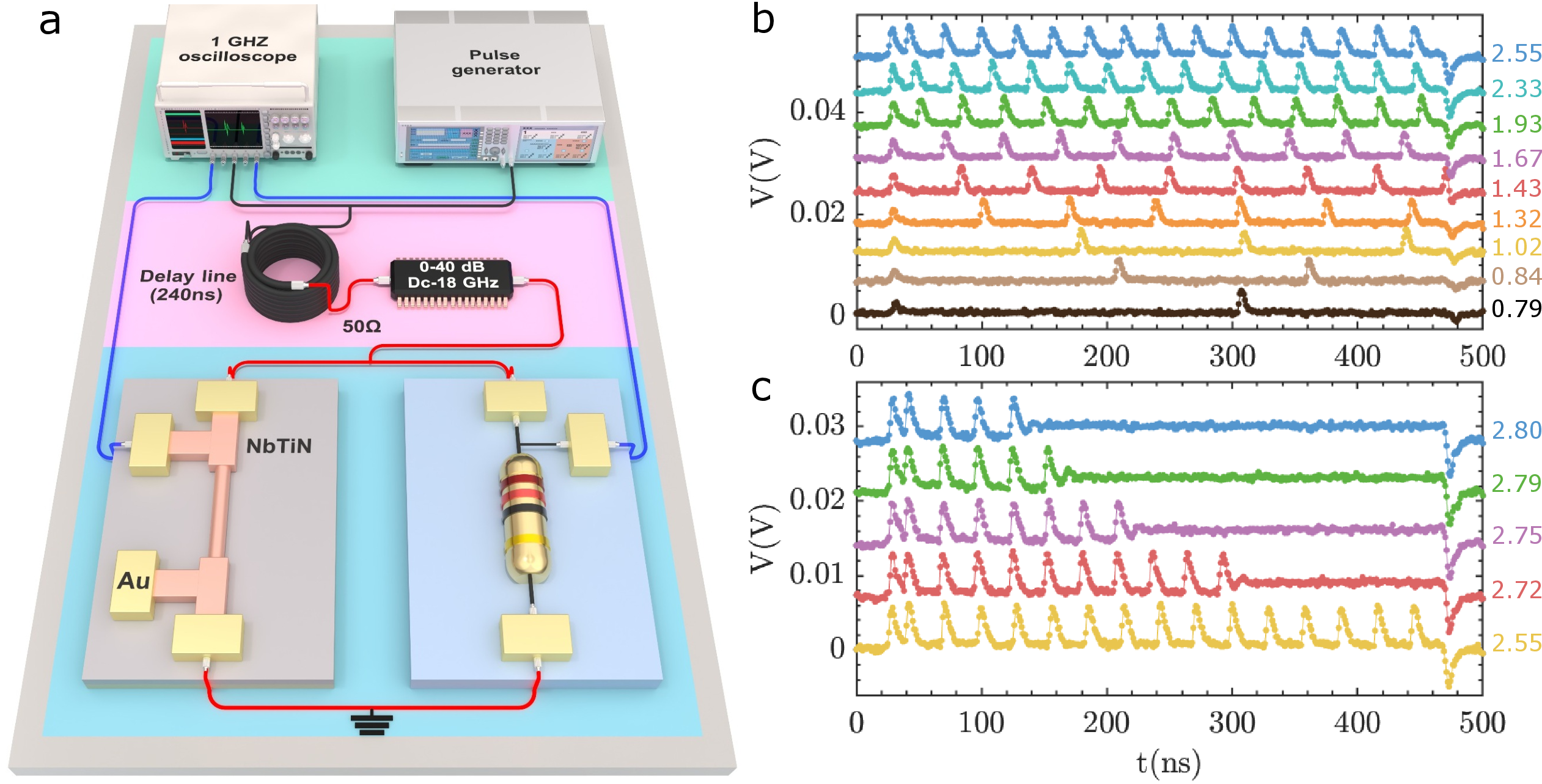}
    \caption{\textbf{Experimental characterization of a spiking superconducting filament.} Transport measurement on a 3$\mu$m-wide NbTiN filament shunted by 1$\Omega$ resistor, at $T=8$K. \textbf{a}, The experimental setup. \textbf{b}, The measured spiking behavior of the voltage, for increasing applied current (shown in mA). Curves are vertically displaced for visibility. \textbf{c}, The death of the superconducting neuron, corresponding to Fig.~\ref{fig:fig1}c,d. The behavior seen in panels b,c has been validated at different temperatures, as shown in the Extended Data Fig.~1.}
    \label{fig:fig2}
\end{figure*}

For the experimental realization of a superconducting neuron, we considered 20~nm thin NbTiN whiskers, measuring 3$\mu$m in width. A metallic resistor was mounted in parallel with the superconducting wire, with resistance values of 0.3 - 2$\Omega$. The filaments were biased with a current pulse of 450~ns duration, in a transport setup depicted in Fig.~\ref{fig:fig2}a. The measurements were performed in a closed-cycle cryostat, to maintain a stable temperature of 5-8K, for a critical temperature of about 10K. For other details, we refer the reader to the Methods section.

Figure~\ref{fig:fig2}b shows the measured temporal evolution of the voltage along the wirelet shunted by 1$\Omega$, for an applied current slightly exceeding the critical current $I_c$ and temperature 8K. As predicted by theory, the delay time decreases and the spiking frequency increases as the bias current is raised, reflecting the enhanced rate of hot-spot formation and recovery in the resistive regime. This tunability of the oscillation frequency through the bias current provides a natural control parameter for neuron-like operation, analogous to the firing-rate modulation in biological neurons. By adjusting the bias level, one can continuously vary the spike rate, amplitude, and temporal regularity, offering a direct means to encode analog information in the spiking frequency. Such current-controlled frequency modulation is thus a convenient mechanism for input processing, enabling electronic adaptation of the neuron’s response with respect to the posed input current. However, for the applied current exceeding the onset of phase III in Fig.~\ref{fig:fig1}a (at current $I_d$), the spiking oscillations are gradually giving way to the phase-slip state, as seen in Fig.~\ref{fig:fig2}c. This marks an effective death of the superconducting wirelet neuron in the experiment, as corroborated by theoretical expectations (Fig.~\ref{fig:fig1}c,d).

\begin{figure*}[!t]
    \centering
    \includegraphics[width=\linewidth]{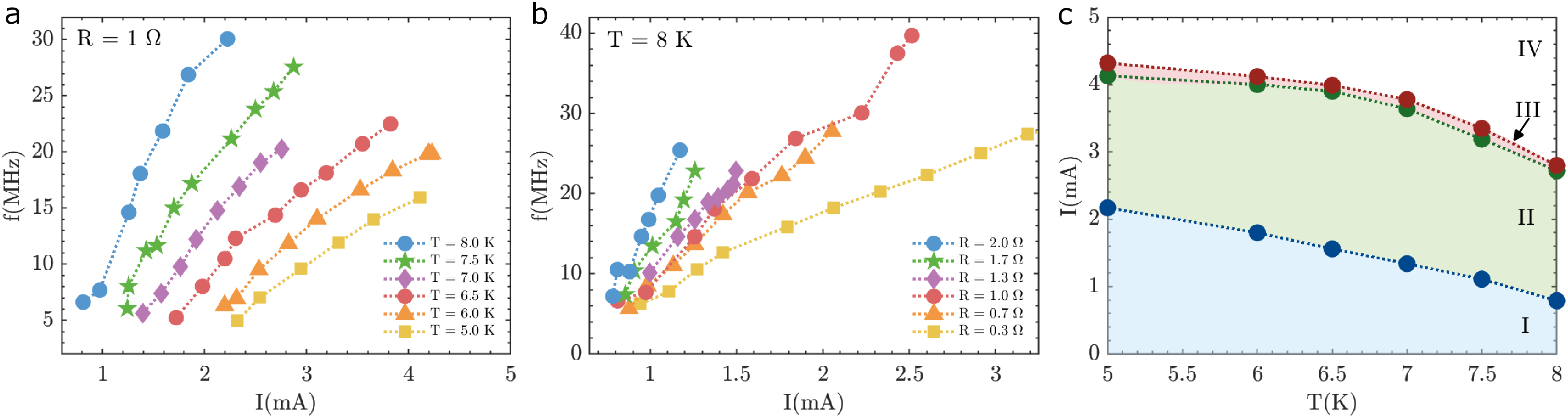}
    \caption{\textbf{Control of spiking frequency.} \textbf{a}, Spiking frequency of the superconducting filament from Fig.~\ref{fig:fig2}, increasing with applied current, for different temperatures. \textbf{b}, Spiking frequency of the same filament, as a function of applied current, at temperature $T=8$~K, but for varied shunt resistance. The spiking behavior has been validated for all used shunt resistances, as shown in Extended Data Fig.~2. \textbf{c}, The full temperature-current phase diagram of the superconducting neuron from Fig.~\ref{fig:fig2}. Phase II is the operational range of the neuron.}
    \label{fig:fig3}
\end{figure*}

For this specific realization of the neuron, i.e. the chosen material, wirelet width and thickness, Figure~\ref{fig:fig3} summarizes the measured spiking characteristics of the superconducting wirelet neuron across a range of experimentally accessible parameters. Panels (a) and (b) show the spiking frequency as a function of the applied bias current for various temperatures and shunt resistances, respectively. The data reveal a highly tunable response, with oscillation frequencies spanning the 4–40 MHz range for bias currents between 0.5 mA and 5 mA. The strong dependence on temperature reflects the interplay between thermal relaxation and critical current modulation, while variations with shunt resistance illustrate how the electrodynamic coupling and energy dissipation govern the oscillatory stability. The corresponding temperature–current phase diagram in Fig.~\ref{fig:fig3}c maps the operational regimes for a wirelet shunted by 1$\Omega$, delineating the active spiking region (green area) that separates the superconducting (I) and normal state (IV). This diagram illustrates a robust experimental window within which neuron-like spiking occurs, demonstrating that both frequency and stability can be precisely engineered via simple tuning of circuit parameters - a valuable feature for further integration in a neuromorphic hardware.

\section*{Pattern recognition with few superconducting neurons}
\begin{figure*}
    \centering
    \includegraphics[width=0.8\linewidth]{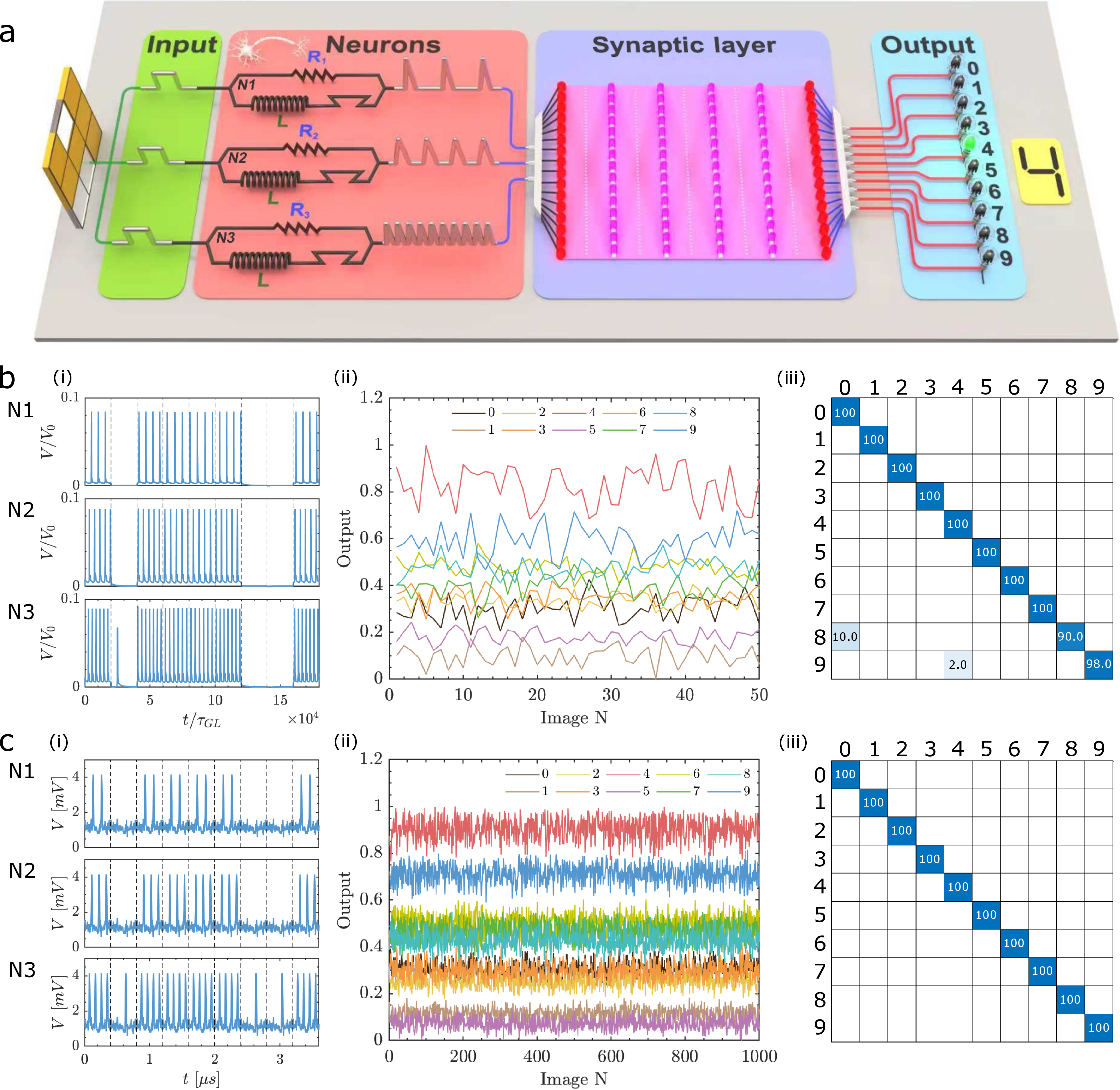}
    \caption{\textbf{Proof-of-concept pattern recognition with three superconducting neurons.} \textbf{a}, Schematic diagram of the circuit, where synaptic multiplexing was software-based. \textbf{b}, Simulated performance of three identical filaments (width $10\xi$, and shunt resistances 0.5, 0.75, and 1$R_{GL}$) in a neural network: (i) exemplified output voltages of each neuron corresponding to the same train of input current pulses for pixels 1-9. (ii) Output result after training, for 50 different images of digit 4. (iii) The input-output matrix for 50 test images of each digit, with percentages of correct and erroneous predictions shown. \textbf{c}, Three experimental NbTiN filaments as in Fig.~\ref{fig:fig2} (width $3\mu$m, and shunt resistances 0.3, 1, and 1.7$\Omega$ top to bottom) in a neural network: (i) exemplified output voltages of each neuron corresponding to the same train of input current pulses for pixels 1-9. (ii) Output result after training, for 1000 different test images of digit 4. (iii) The input-output prediction matrix for 1000 test images of each digit, showing 100\% accuracy.}
    \label{fig:fig4}
\end{figure*}

To demonstrate the integration of individual superconducting wirelets into a functional neuromorphic hardware element, we implement a minimal neural network comprising just three superconducting neurons tasked with recognizing digits (0–9) from $3 \times 3$-pixel digital images, as illustrated in Fig.~\ref{fig:fig4}a. Despite the small number of neurons, the classification task is made non-trivial by introducing graded pixel brightness, thereby increasing the analog complexity of the input. With the maximum brightness normalized to unity, bright pixels take intensity values of 0.8, 0.9, and 1.0, whereas dark pixels correspond to 0.0, 0.1, and 0.2, allowing for subtle contrast variations. Each pixel intensity is converted into a current pulse with an amplitude proportional to its brightness, thus encoding the visual information in the temporal and amplitude domain. These current pulses are then sequentially applied (in a sequence from the top-left to the bottom-right pixel) to the three superconducting neuron elements, which process the input stream in real time and generate a distinct voltage response characteristic of the input pattern. Note that the same train of pulses with same values of current is fed to all three neurons, hence the pixel intensity range has to be converted to a range of current between the maximal $I_c$ and the minimal $I_d$ of the three neurons.

The same pattern recognition problem was then solved using both theoretical simulation and the experimental data, for cross-reference. The theoretical results are shown in panels within Fig.~\ref{fig:fig4}b. Panel \textbf{b}(i) shows the calculated voltage response of each neuron (shunt resistance increases from top to bottom) for the exemplified case of digit $4$. The black dashed lines delimit the current pulses associated with each pixel (see Supplementary Video 2 for the order parameter dynamics). The voltage responses of three neurons are then combined into a single array $V(t) = [V_1(t), V_2(t), V_3(t)]$. In the synaptic processing (see Methods for details), the array $V(t)$ is linearly transformed using the cross-entropy loss function method\cite{Goodfellow2016} to obtain a $1 \times 10$ vector, where each element gives the probability that the input image corresponds to each digit 0-9. These elements are exemplified in panel \textbf{b}(ii) for digit $4$ as input. The final output, i.e. the prediction of the neural network, is taken to be the digit corresponding to the maximum probability. 

In the simulations, the linear transformation process was trained on a set of $250$ images of each digit. We have subsequently tested the accuracy of the network using a set of $50$ images of each digit, that were not used in training. The results are presented in panel \textbf{b}(iii), as a diagram of the prediction distribution for each digit. Although the considered architecture was minimalistic, the prediction was $100\%$ accurate for digits from $0$ to $7$, $90\%$ for $8$ and $98\%$ for digit $9$. For the same network built with the experimental data, we show the results in Fig.~\ref{fig:fig4}c. Identical procedure was followed, but for both training and test sets increased to $1000$ images of each digit. As a result, as shown in panel \textbf{c}(iii), we have obtained the \textit{$100\%$ prediction accuracy} for all the input images considered.

\begin{figure*}
    \centering
    \includegraphics[width=\linewidth]{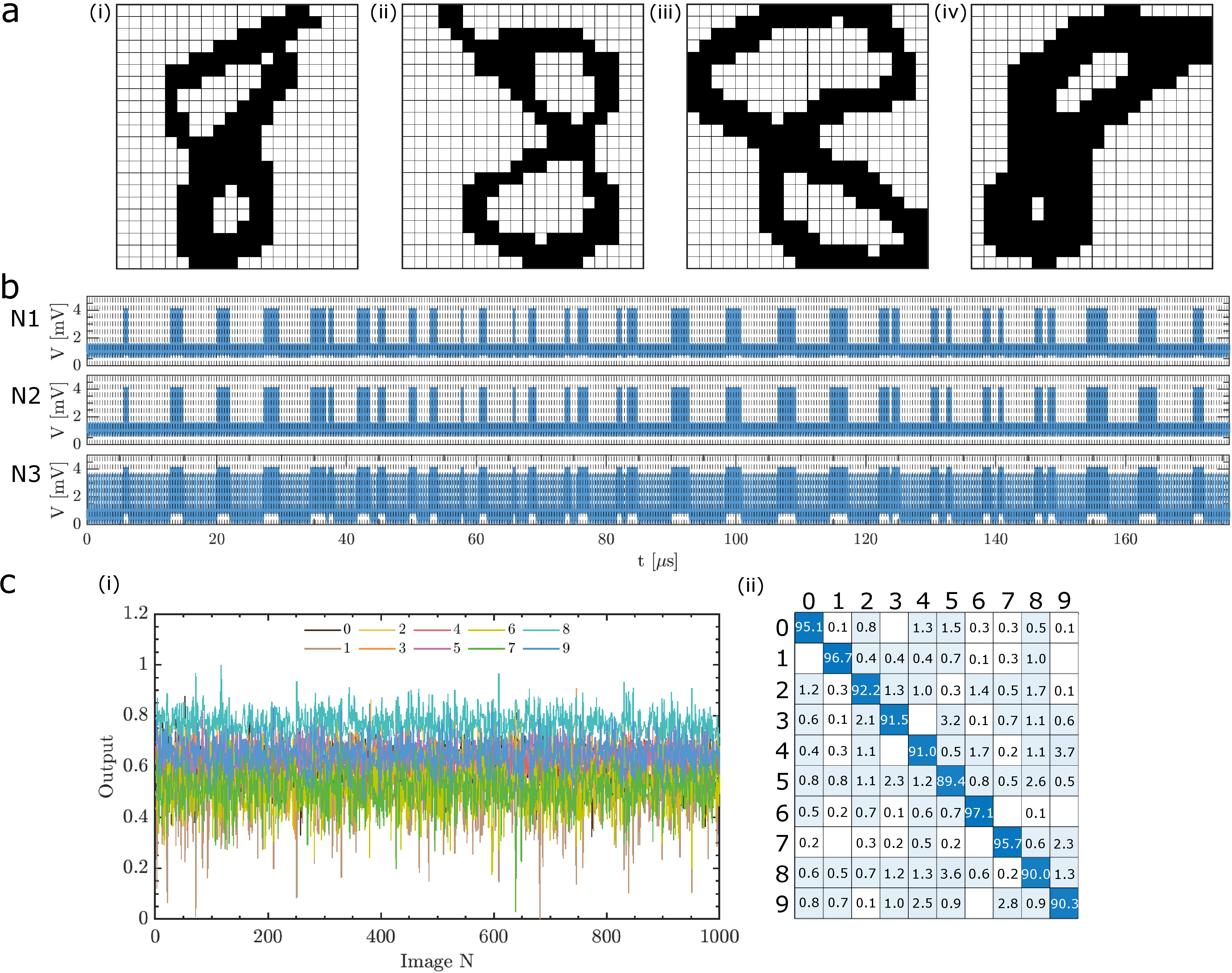}
    \caption{\textbf{A complex task realized with same three-neuron network.} \textbf{a}, Illustrative examples of hand-written number 8 from the MNIST database. \textbf{b}, The temporal voltage output of the three neurons for the input sequence of 420 image pixels of the image a(i). \textbf{c}, Output result after synaptic processing of 1000 test images of digit 8. \textbf{d}, The full input-output matrix on 1000 test images of each digit, demonstrating 92.9\% accuracy on average.}
    \label{fig:fig5}
\end{figure*}

To further evaluate the capability and scalability of our minimal superconducting neural architecture, we increased the task complexity from simple $3 \times 3$ patterns to pixelated representations of handwritten digits with a resolution of $22 \times 20$ pixels, available from the MNIST database\cite{LeCun1998}. Each pixel was again encoded as a current pulse and sequentially fed to the same three superconducting neuron elements as used in Fig.~\ref{fig:fig4}, preserving the same circuit architecture as in the simpler task. Despite the dramatic increase in input dimensionality and image variability, the system was successfully trained on $3000$ images per digit and tested on $1000$ unseen examples, as shown in Fig.~\ref{fig:fig5}. Remarkably, the resulting average prediction accuracy reached 92.9\%, demonstrating that even such a compact superconducting wirelet network (only three neurons) can extract and classify complex spatial patterns (see Supplementary Video 3 for the evolution of average and minimum accuracy with training epochs). From the standpoint of neural network theory, this performance illustrates the strong nonlinear feature mapping intrinsic to our superconducting neurons. Moreover, according to established scaling principles in deep learning\cite{Goodfellow2016,LeCun2015,Hestness2017}, increasing the number of input neurons is expected to rapidly boost the classification accuracy above best-performing digital neural systems. These findings reaffirm the scalability and promise of superconducting neuromorphic circuits for high-speed, low-power pattern recognition at large data scales\cite{Shainline2019,Chaudhary2023}.

\subsection{On-chip training with superconducting wirelet neurons}
In our results, synaptic training of the experimental neuron responses was performed using designated software. In fact, there is a logical extension of the circuitry depicted in Fig.~\ref{fig:fig4}a that enables full hardware on-chip training of the neural network. In this pattern-recognition circuit, an array of superconducting wirelet neurons (as many as needed, distinctly shunted) receive the time-encoded input current chain of pulses representing the pixel intensities of an image. The resulting voltage spikes from all input neurons are then combined sequentially in time to form an aggregate output waveform, which is subsequently converted into a current pulse and distributed among multiple superconducting output wirelets (each representing different possible outcomes of the operation), not necessarily identical to the input ones. Unlike the input stage, these output wirelets are not shunted but instead incorporate electrostatic gates that modulate their local critical current\cite{Giazotto2020,DiBernardo2020} and, consequently, the partitioning of the output current from the input stage. Such modulation by gating is a convenient and robust mechanism for signal processing and training, enabling dynamic adaptation of the response without requiring any structural reconfiguration of the superconducting circuit. In this architecture, the gate voltages serve as trainable synaptic weights. Note that the gate voltage is time dependent in this architecture, as its distributed weight over the entire input waveform needs to be trained. By sequentially feeding the voltage responses of all input neurons and applying a time-dependent gating function, each output wire effectively performs a weighted summation of the entire input sequence in real time. During training, the gate voltages are tuned such that the wirelet producing the largest output voltage corresponds to the correct outcome. In the task of Fig.~\ref{fig:fig5}, ten output wirelets would be required, each corresponding to different recognized digit (0–9). 

Such an implementation of the gated superconducting synaptic stage thus enables compact, yet energy-efficient and reconfigurable synaptic functionality within a fully cryogenic computing platform. This architecture offers a key advantage in that the training and inference stages are inherently unified: the same superconducting circuitry responsible for generating the output response also enables direct electronic reconfiguration of the synaptic weights, allowing rapid adaptation to different recognition tasks without external hardware modification. All functional elements, be it neurons, interconnects, and gated synapses, can be fabricated on a single superconducting chip, ensuring scalability and low parasitic losses. When implemented in niobium-based technology, the system is fully compatible with established superconducting electronic and quantum-circuit platforms, promising seamless integration with RSFQ logic, cryogenic memory, and qubit architectures. This combination of reconfigurability, compactness, and technological compatibility positions our approach as a promising route toward scalable, energy-efficient neuromorphic processors operating entirely within the superconducting domain.

\section*{Discussion and conclusions}

\subsection{Energy efficiency}
In our superconducting wirelet neuron architecture, the total energy associated with a spiking operation can be decomposed into three primary contributions. The spike-generation energy, corresponding to the formation of a voltage pulse by the firing nanowire, is approximately 100~fJ. The subsequent voltage-to-current conversion stage dominates the overall budget, requiring about 1 pJ per spike to drive the output network. The resulting current is distributed among ten output wirelets, each functioning as a gated (weighted) synapse that converts the received current back into a voltage pulse, consuming up to 100~fJ per output event. When this total dynamic energy is amortized across the ten synaptic outputs, the system-level energy per synaptic event lies well below 0.5~pJ, positioning this platform among the most energy-efficient neuromorphic hardware implementations reported to date.

For comparison, state-of-the-art digital neuromorphic processors such as ODIN~\cite{Frenkel2019ODIN} and Intel Loihi-2~\cite{Davies2021Loihi2} demonstrate minimum energies of approximately 10–20~pJ per synaptic operation, while analog mixed-signal systems such as BrainScaleS-2~\cite{Pehle2022BrainScaleS2} achieve similar magnitudes. Photonic neuromorphic platforms, in contrast, have recently achieved femtojoule-level pulse energies\cite{Lee2022PhotonicSpiking}, and sub-picojoule energy per synaptic event\cite{Zeng2025PhotonicDFB}, albeit with large hardware overheads (lasers, drivers, electronics). Our superconducting wirelet approach therefore bridges the gap between low-energy yet experimentally demanding photonic spiking schemes and scalable electronic implementations, offering sub-picojoule system-level synaptic event energy with intrinsic cryogenic compatibility. Furthermore, the energy per spike and energy per operation in our system can still be significantly reduced, since the operating current (and thus the switching energy) scales with both the wirelet width and the critical current density of the superconducting material. Our present devices employ 3$\mu$m-wide NbTiN wirelets, leaving substantial headroom for improvement through geometric scaling and material optimization, potentially pushing energy efficiencies well into the deep-femtojoule regime.

\subsection{$\mu$-neuron vs. nano-neuron}
Although the concept of wirelet-based superconducting neurons does not strictly depend on the wirelet size, the choice between micrometer- and nanometer-scale superconducting wires is important in defining the performance and reliability of neuromorphic circuits. Micrometer-wide wires, as used in our experiment, offer enhanced stability, thermal robustness, and signal strength, making them well suited for reproducible spiking behavior. Their larger cross-section supports higher critical currents and more effective heat dissipation, reducing susceptibility to quenching and thermal noise during high-frequency operation\cite{Zhou2019,Berggren2018}. In addition, the greater volume accommodates stronger pinning, yielding cleaner, more predictable electrical responses in presence of any stray magnetic field and vortices\cite{Cheng2017}. These attributes collectively enable robust and biologically faithful spiking dynamics, critical for system-level reliability. Nanowires, by contrast, offer obvious advantages in scaling and energy efficiency, enabling dense interconnects and lower operating energies, but at the cost of reduced thermal stability and increased fabrication and integration complexity\cite{Shainline2019}. Consequently, while nanowires remain attractive for ultra-compact, low-power architectures, micrometer-scale implementations currently provide a more balanced platform, while still leaving room for energy and footprint improvements through geometric downscaling.

In summary of the above results and discussion, we have established shunted superconducting wirelets as a minimal, yet fully functional, artificial neuron platform, capable of spike generation and firing behavior tunable through current bias, thermal control and engineered shunt resistance. We performed synaptic training by harnessing the temporal waveform of the voltage response of a wirelet to a sequence of input current pulses, to demonstrate pattern recognition of handwritten digits with only three neurons - an extreme reduction in architectural complexity compared with existing artificial neural systems, yet achieving outstanding classification fidelity. This approach also enables full on-chip learning, in which gated output wirelets are introduced and the synaptic training is performed directly on them, such that the synaptic and output layers are integrated. This design eliminates the need for a separate, complex synaptic network and achieves learning and inference within the same hardware layer. Altogether, our concept outlines a scalable route toward ultralow-energy neuromorphic hardware, readily integrable with emerging superconducting and quantum electronic technologies, and marks a frontier in superconducting artificial intelligence, where architectural simplicity meets high efficiency and functionality.

\begin{methods}
\paragraph{Experimental setup}
The experimental configuration followed and extended methodologies used in prior studies of current-biased superconducting nanowires under pulsed excitation\cite{Harrabi2019}, while adapted here for the micrometer-scale NbTiN wirelets and external tunable shunts.

Thin NbTiN superconducting whiskers, 20~nm thick and 3~$\mu$m wide, were fabricated on sapphire and SiO$_2$/Si substrates at STARCryoelectronics (New Mexico, USA). Deposition was carried out under ultra-high-vacuum conditions to ensure optimal film uniformity and low defect density. The patterning of the superconducting filaments and the definition of gold electrical contacts were performed using optical photolithography followed by reactive ion etching and argon ion milling to achieve clean superconducting edges and reliable electrical interfaces. The final device layout, shown schematically in Fig.~\ref{fig:fig2}a, features a central superconducting stripe approximately 800~$\mu$m long, connected at both ends to gold pads for electrical access.

To investigate the spiking characteristics of the superconducting wirelets, each device was connected in parallel with an external metallic resistor mounted off-chip. The shunt resistor values ranged from 0.3~$\Omega$ to 2.0~$\Omega$ (specifically 0.3, 0.7, 1.0, 1.3, 1.7, and 2.0~$\Omega$), allowing systematic control of the current redistribution and damping. One end of the superconducting stripe was connected to a voltage-pulse source, while the opposite end was grounded together with the shunt resistor. This arrangement provided a simple yet effective test platform for studying the interplay between shunting and superconducting spiking dynamics. To verify that the observed spiking behavior was not specific to a particular device geometry, additional samples with stripe widths of 5~$\mu$m and 10~$\mu$m were fabricated and tested under identical experimental conditions. The characteristic spiking dynamics and delay-time dependencies were consistently reproduced across all wire widths, confirming the robustness and generality of the observed phenomena.

All measurements were performed in a vacuum environment within a closed-cycle cryostat, enabling stable temperature control above 5~K. The NbTiN whiskers exhibited a superconducting transition temperature ($T_c$) of approximately 10~K, consistent with high-quality films of similar stoichiometry.

Voltage pulses of 450~ns duration and 10~kHz repetition rate were applied to the devices, with pulse amplitude varied to explore operation above and below the critical current. The waveform was delivered through an air-delay transmission line with a propagation delay of 240~ns, ensuring well-timed and reproducible excitation conditions. Owing to the 50~$\Omega$ coaxial cable impedance, the experimental configuration operated in an effectively current-biased regime.

The temporal voltage response was measured simultaneously across both the superconducting wire and the parallel metallic shunt using a high-speed digital oscilloscope with sub-nanosecond temporal resolution and 50~$\Omega$ input termination. Each dataset consisted of time-domain traces averaged over multiple pulse cycles to minimize noise and improve signal-to-noise ratio. The measured voltage–time profiles were then analyzed to extract spiking delay times, frequency responses, and the dependence on applied current and temperature, as discussed in the main text.

\paragraph{Time-dependent Ginzburg-Landau simulations}
The numerical modeling of each shunted superconductor was done by solving the time-dependent Ginzburg-Landau equation:
\begin{equation}
    \frac{u}{\sqrt{1+\gamma^2|\psi|^2}}\left ( \frac{\partial }{\partial t}
    +\text{i}\varphi + \frac{\gamma^2}{2}\frac{\partial |\psi|^2}{\partial t^2} \right ) \psi = \left(\mbox{\bm $\nabla$}-i\textbf{A}\right)^2\psi
    +\psi(1-|\psi|^2),
    \label{eq:eq1}
\end{equation}
where the order parameter $\psi$ is scaled to its value in absence of magnetic field $\psi_0$; all lengths are expressed in units of the superconducting coherence length $\xi$; field and vector potential are in units of $H_{c2}$ and $H_{c2}\xi$, respectively, with $H_{c2}$ being the bulk upper critical field $\Phi_0/2\pi\xi^2$; time is in units of $t_{GL}=\pi\hbar/8uk_BT_c$, with $u = 5.79$ being a constant coming from microscopic theory. The parameter $\gamma$ is the product of inelastic electron-phonon scattering time and the GL gap at $T = 0$, $\Delta_{GL}$. Without loss of generality, we used $\gamma = 10$ in our calculations. Finally, the scalar potential $\varphi$ is given in units of $V_{GL}\hbar/2et_{GL}$. At each time step, the scalar potential is obtained by solving the Poisson equation:
\begin{equation}
    \bm{\nabla}^2\varphi = \bm{\nabla}(\textrm{Im}\left [\bar{\psi}(\bm\nabla-i\textbf{A})\psi\right]).
    \label{eq:eq2}
\end{equation}
Eqs.~\ref{eq:eq1}-\ref{eq:eq2} are then numerically solved (see Ref.~\citenum{milovsevic2010} for details on the discretization), using a grid size $\Delta x= \Delta y=0.2\xi$. The current density $J_s$ transported by the superconducting wirelet of size $L\times w$ is introduced through the boundary conditions for the scalar potential as $\bm{\nabla}\varphi(\pm L/2,y) = J_s\bm{\hat{x}}$. At the remaining boundaries of the wirelet we have $\bm{\nabla}\varphi(x,\pm w/2) = 0$. For the order parameter, we have $\psi(\pm L/2,y) = 0$ at the current leads, and $(\bm\nabla-i\textbf{A})\psi|_y = 0$ at $y = \pm w/2$, assuring no supercurrent flows out of the sample on those boundaries.

To incorporate the effects of the shunt, we simultaneously solved for the time evolution of the transport current density in a shunted electrical system (see inset of Fig.~\ref{fig:fig1}a), from equation
\begin{equation}
    L_K\frac{dJ_s}{dt} = (J-J_s)R_{sh}-V,
    \label{eq:eq3}
\end{equation}
where $J$ is the total current density applied to the circuit; $L_K$ is the circuit kinetic inductance in units of $L_{GL} = (\Phi_0e^*\pi/8k_bu)/(\sigma_n\xi T_c)$, with $\sigma_n$ being the normal state conductivity of the superconductor material; $R_s$ is the shunt resistance, in units of $R_{GL} = 1/(\sigma_n\xi)$; and $V$ is the voltage drop across the superconducting wire. At each time step, Eq.~\ref{eq:eq3} is solved and the current density $J_s$ carried by the superconductor is recalculated.

\paragraph{Synaptic training}
The training of the neural network, realized in the synaptic layer of the circuit, was carried out using a dedicated control routine made into a custom-tailored software. In this process, the time-dependent voltage output of each of the three neurons $V_i(t)$ was discretized in $N$ steps equidistant in time, such that the cumulative waveform vector of those discretized arrays $\mathcal{V} = [V_1^d, V_2^d, V_3^d]$ has $3N$ elements. For the $3 \times 3$ images discussed in Fig.~\ref{fig:fig4}, we had $N = 1000$ for the training using numerical simulation results and $N = 1278$ for the training using experimental results. When working with the MNIST database, $N$ was set to $62480$.

The output of the network, a $10$-element array $y$ (one element for each possible digit) was then obtained by a linear transformation of $\mathcal{V}$:
\begin{equation}
    y = \mathcal{V}W
\end{equation}
where the weights are contained in the $3N \times 10$ matrix $W$. The targets $T$ are then one-hot vectors with $10$ elements.

For each epoch $n_{eph}$ of the training process, we compute the output activations using a \textit{softmax} function:
\begin{equation}
    s = \frac{\text{exp}(y)}{\sum_{i=1}^{10}\text{exp}(y_i)}.
\end{equation}
The cross-entropy loss function was then employed to quantify the prediction error between the network output and the target label:
\begin{equation}
    L_{n_{eph}} = -\sum_{i=1}^{10} T \cdot \text{log}(s).
\end{equation}
Analytically, one can define the gradient of the cross-entropy loss with respect to the weights as:
\begin{equation}
    G = \mathcal{V}^T(s-T).
\end{equation}
The weight matrix parameters were updated via stochastic gradient descent (SGD) according to:
\begin{equation}
    W = W-\eta G,
\end{equation}
with $\eta = 0.01$ being the learning rate. The same procedure was reiterated through successive training epochs, allowing progressive adjustment of the synaptic weights until the network reached convergence.
\end{methods}

\bibliography{bibliography}

\begin{addendum}
 \item [Acknowledgments] K.H. gratefully acknowledges the support of the King Fahd University of Petroleum and Minerals, Saudi Arabia, under project ISP23212. L.R.C. and M.V.M. acknowledge support from the Research Foundation-Flanders (FWO-Vlaanderen).
 \item [Author contributions] Conceptualization: M.V.M. Methodology: K.H., L.R.C. and M.V.M. Investigation: All authors. Visualization: K.H. and L.R.C. Funding acquisition, project administration and supervision: K.H. and M.V.M. Writing, review and editing: All authors.
 \item[Competing Interests] The authors declare no competing interests.
 \item[Correspondence] Correspondence and requests for materials should be addressed to Khalil Harrabi (email: harrabi@kfupm.edu.sa) and Milorad V. Milo\v{s}evi\'c (email: milorad.milosevic@uantwerpen.be).
\end{addendum}

\subsection{Extended Data}
\renewcommand{\thefigure}{ED\arabic{figure}}
\setcounter{figure}{0}

\begin{figure*}[!b]
    \centering
    \includegraphics[width=\linewidth]{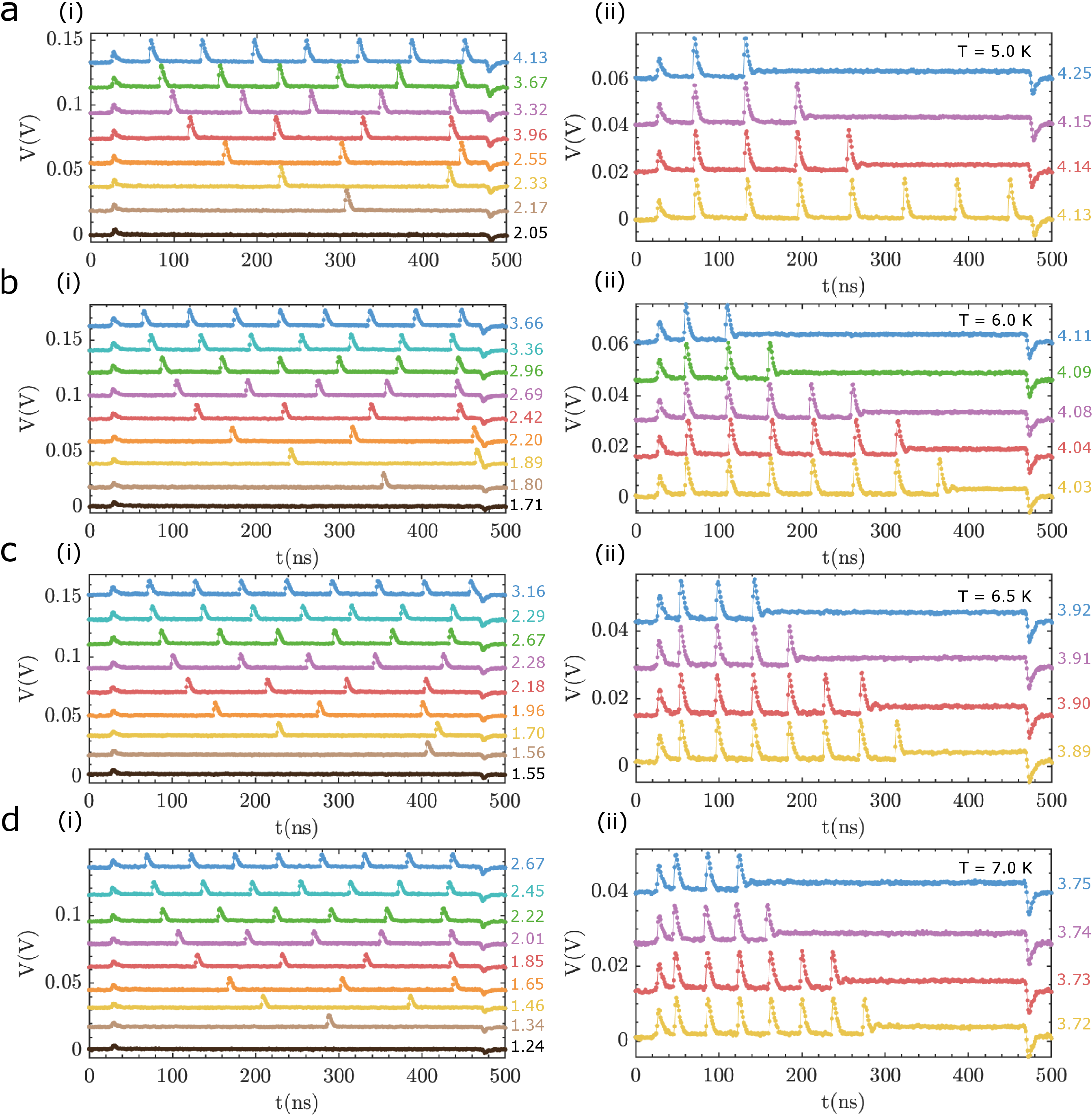}
    \caption{\textbf{Experimental characterization of a spiking superconducting filament at different temperatures.} Transport
measurement on a 3$\mu$m-wide NbTiN filament shunted by 1$\Omega$ resistor, analogous to Fig.~\ref{fig:fig2}\textbf{b}, and \textbf{c}, of the main text, but at different temperatures (indicated in the figure). Panels $(i)$ and $(ii)$ show the spiking behavior of the voltage and the death of the superconducting neuron. Currents are given in mA.}
    \label{fig:fig1SI}
\end{figure*}

\begin{figure*}[!b]
    \centering
    \includegraphics[width=\linewidth]{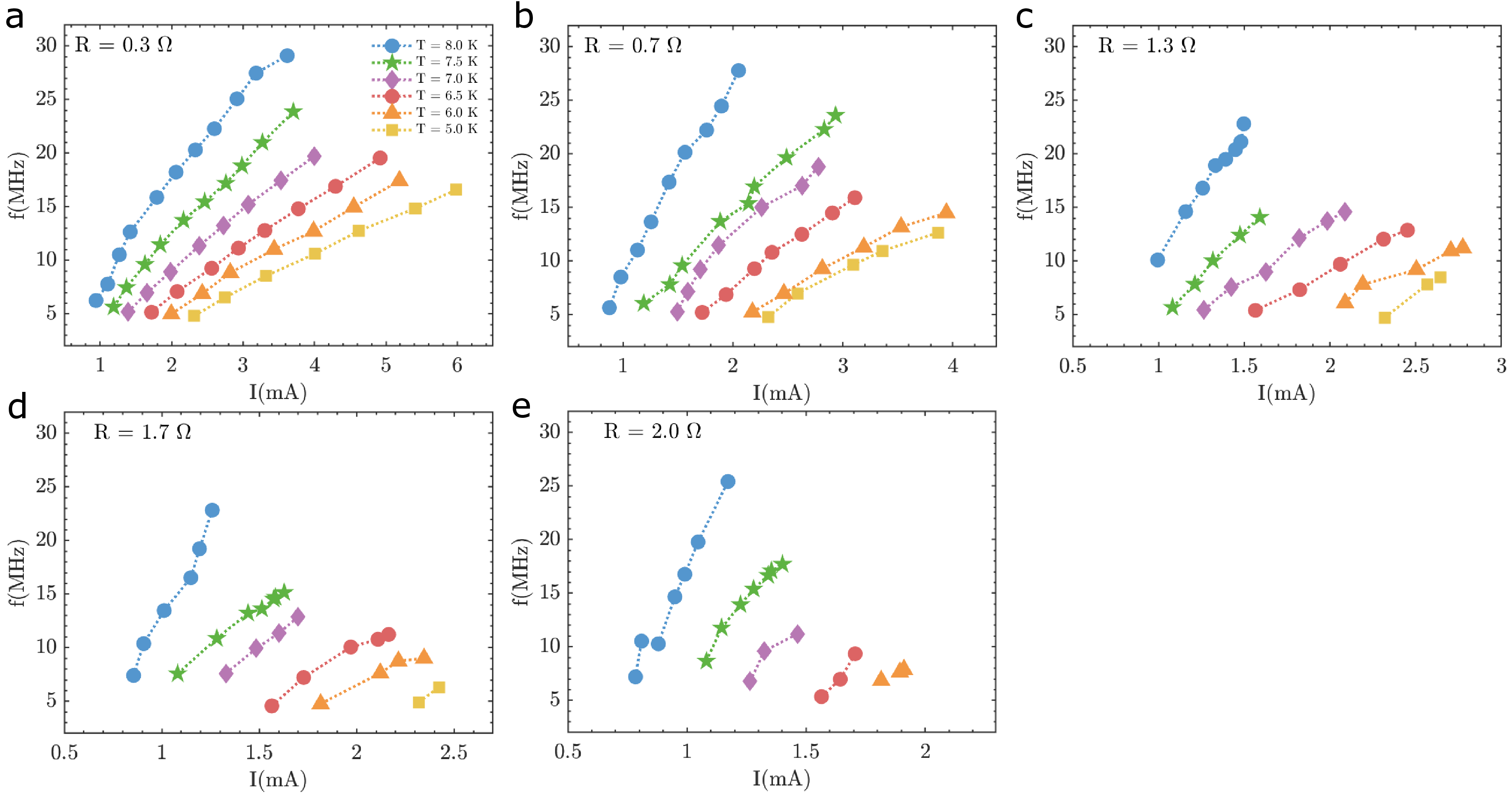}
    \caption{\textbf{Spiking frequency for different shunt resistance.} The spiking frequency as a function of applied current at different temperatures. Shown panels are analogous to Fig.~\ref{fig:fig3}a of the main text, but correspond to different shunt resistances used, as indicated in the figure.}
    \label{fig:fig2SI}
\end{figure*}

\end{document}